# 2D materials and van der Waals heterostructures

*K. S. Novoselov[1,2]\*, A. Mishchenko[1,2], A. Carvalho[3], A. H. Castro Neto[3]\**

[1]*School of Physics & Astronomy, University of Manchester, Oxford Road, Manchester, M13 9PL, UK*
[2]*National Graphene Institute, University of Manchester, Manchester, M13 9PL, UK*
[3]*Centre for Advanced 2D Materials and Graphene Research Centre, National University of Singapore, 2 Science Drive 3, Singapore, 117542*

*The physics of two-dimensional (2D) materials and heterostructures based on such crystals has been developing extremely fast. With new 2D materials, truly 2D physics has started to appear (e.g. absence of long-range order, 2D excitons, commensurate-incommensurate transition, etc). Novel heterostructure devices are also starting to appear - tunneling transistors, resonant tunneling diodes, light emitting diodes, etc. Composed from individual 2D crystals, such devices utilize the properties of those crystals to create functionalities that are not accessible to us in other heterostructures. We review the properties of novel 2D crystals and how their properties are used in new heterostructure devices.*

The family of 2D materials(*1*) has grown appreciably since the discovery of graphene(*2*). Each new material brings excitement and puzzles as their properties are usually very different from their 3D counterparts. Furthermore, 2D materials offer a huge flexibility in tuning of their electronic properties. Thus, band gap engineering can be done by changing the number of layers(*3*, *4*). Even more interesting is the specific 2D physics observed in such materials (for instance Kosterlitz-Thouless (KT) behavior, characterized by the emergence of topological order, resulting from the pairing of vortices and antivortices below a critical temperature). Crystals with transition metals in their chemical composition are particularly prone to many-body instabilities such as superconductivity, charge (CDW) and spin (SDW) density waves. Such effects can also be induced by proximity if such crystals are sandwiched with other 2D materials.

Heterostructures of 2D materials offer not only a way to study these phenomena, but open unprecedented possibilities of combining them for technological use. Such stacks are very different from the traditional 3D semiconductor heterostructures, as each layer acts simultaneously as the bulk material and the interface, reducing the amount of charge displacement within each layer. Still, the charge transfers between the layers can be very large, inducing large electric fields and offering interesting possibilities in band-structure engineering.

Among the tools for band-structure engineering in the van der Waals heterostructures are the relative alignment between the neighboring crystals, surface reconstruction, charge transfer and proximity effects (when one material can borrow the property of another by contact via quantum tunneling or by Coulomb interactions). Thus moiré structure for graphene on hexagonal boron nitride (hBN) leads to the formation of the secondary Dirac points(*5-9*), commensurate-incommensurate transition in the same system leads to surface reconstruction(*10*) and gap opening in the electron spectrum(*8*), spin-orbit can be induced in graphene by neighboring transition metal dichalcogenide (TMDC)(*11*, *12*).

We provide a review of 2D materials, analyzing the physics that can be observed in such crystals. We discuss how these properties are put to use in new heterostructure devices.

## Transition metal dichalcogenides

TMDCs, with formula $MX_2$, where $M$ is a transition metal and $X$ a chalcogen, offer a broad range of electronic properties from insulating or semiconducting (e.g. Ti, Hf, Zr, Mo and W dichalcogenides) to metals or semimetals (V, Nb and Ta dichalcogenides). The different electronic behavior arises from the progressive filling of the non-bonding $d$ bands by the transition metal electrons. The evolution of the electronic density of states (DoS) is shown on Fig. 1 (adapted from Refs(*13-17*)) for the most stable phase of each of the dichalcogenides.

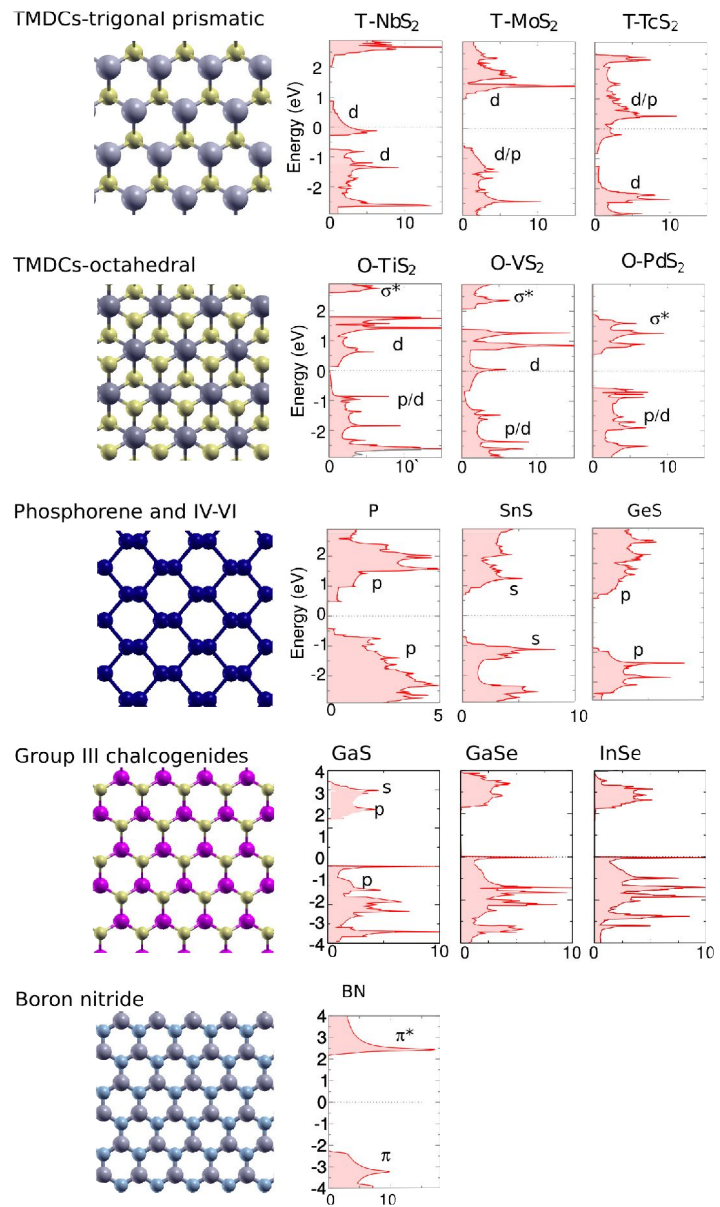

**Fig. 1 Electronic properties of different classes of 2D materials.** The Fermi level is set to the zero of the energy scale. The DoS is given in states/eV per cell.

All TMDCs have a hexagonal structure with each monolayer comprising three stacked layers (*X-M-X*). The two most common polytypes of the monolayers are designated trigonal prismatic (e.g. $MoS_2$, $WS_2$) and octahedral (e.g. $TiS_2$), referring to the coordination of the transition metal atom. Inversion symmetry is broken in the former, giving rise to piezoelectricity and having important consequences for the electronic structure. In addition, many of the tellurides, $TcS_2$, $ReS_2$ and other dichalcogenides adopt lower symmetry structures where the metal atom is displaced away from the centre of the coordination unit.

## Metallic TMDC

As one can see in Fig. 1, the DoS of metallic TMDC has two main properties: the Fermi level of the undoped material is always crossing a band with *d*-orbital character, implying that the electrons move mostly in the metal layers; the DoS at the Fermi level is usually quite high which hints to a common explanation for the phase transitions which are observed in these materials([18](#)).

The interest in these materials comes from the existence of CDW and superconductivity in their phase diagram([19](#)). While the CDW phase has clear insulating tendency (opening a gap and suppressing the DoS at the Fermi level) the superconducting phase needs finite DoS in order to exist, leading to a direct competition between the two many-body states. This competition leads to a complex phase diagram with the presence of inhomogeneous electronic and structural patterns, which have been observed in electron microscopy and neutron scattering in the 3D parent compound. Measurements of specific heat and magnetic susceptibility in 3D samples have shown partial gapping of the Fermi surface. In some cases ($TaS_2$) the CDW transition leads to the decoupling of the unit cells along the axis perpendicular to the planes with an enormous increase of the transverse resistivity.

These unusual properties of metal TMDCs have been the subject of intense theoretical debate but no consensus has been reached. The mechanism for the CDW transition does not fit standard weak coupling mean field theories such as Fermi surface nesting or transitions induced by van Hove singularities. Many angle resolved photoemission experiments have been performed in 3D samples with contradictory results([20](#)). The existence of several Fermi surface sheets and the partial gapping of the Fermi surface make the theoretical interpretation of the experimental data quite difficult. Furthermore, the co-existence of CDW and superconductivity (clearly seen in local probes)([21](#)), indicates that many-body effects play a very important role in these materials.

Important information can be obtained from transport data in these materials when done in conjunction with the application of electric and magnetic fields. External electric field changes the Fermi energy and the carrier concentration in the 2D material, without the need of chemical doping (which was the case in 3D materials and which introduces appreciable disorder).

In a recent experiment on 1T-$TiSe_2$ a 2D film was encapsulated by hBN and subject to transverse electric and magnetic fields([22](#)). By applying an external electric field to change the carrier density it was possible to tune the CDW transition temperature from 170 K to 40 K and, concomitantly, the superconducting transition temperature from 0 K to 3 K. The control of the transition temperatures using electric field allows the critical exponents for the phase transition to be determined with high accuracy. Moreover, applying an external transverse magnetic field at the same time revealed novel physical phenomena associated with periodic motion of the Cooper pairs in the superconducting phase. Such behavior seems to be tied up with the formation of discommensurations between different CDW domains, namely, the electronic system broke down in perfectly ordered superconducting and CDW domains.

## Phase transitions in 2D materials

Electrons in a solid are characterized by several quantum numbers that include charge and spin. Due to electron-electron or electron-ion interactions, electrons can organize themselves in phases which are characterized by an order parameter which is associated with these degrees of freedom. In a CDW, as in the case of TMDCs, the order parameter is the local electron density, $\rho(r)$, where $r$ is the position vector, which orders with a well-defined periodicity. This periodicity implies that the Fourier transform of the density, ρ($Q$), where $Q$ is the so-called ordering wave-vector of the CDW, acquires a finite expectation value. For a CDW the expectation value of ρ($Q$) is the order parameter which is zero in the disordered (or normal) phase, and finite in the ordered phase. The transition between these phases can be driven by external forces such as electric, mechanical, thermal, etc.

2D systems play a particular role in the physics of phase transitions. For a system with a continuous order parameter it was shown([23](#)) that it is not possible to have true long range order in dimensions smaller than three at any finite temperature $T$, implying that even minute thermal fluctuations can destroy order. In 2D long range order is only possible at strictly zero temperature. At $T=0$ it is also possible for a system to be disordered if one varies an external parameter such as pressure or electric field, $E$, Fig. 2. The point at which a system becomes ordered at $T=0$ is called quantum critical point and the transitions are called quantum phase transitions. Notice that in this case it is not thermal motion that drives the system from order to disorder but quantum fluctuations. In this type of transition the scale at which order is created is characterized by a correlation length, ξ, which diverges at the phase transition as:

$$\xi(E) \sim 1/|E-E_c|^\nu ,$$

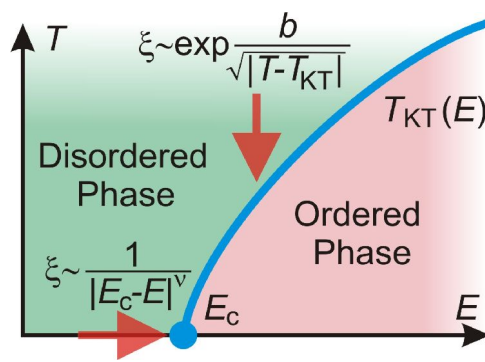

**Fig. 2 Phase diagram for a 2D material with a quantum phase transition.**

where $E_c$ is the critical field and ν is the critical exponent. Fluctuations of the order parameter at different points in space decay exponentially with ξ. Fluctuations in length scales lead to fluctuations in energy scales as well. In a second order phase transition the characteristic energy scale, Δ, associated with the particular order (that is, the energy gap in the system) vanishes at the phase transition with another dynamical exponent, z, as:

$$\Delta(E) \sim 1/\xi^z \sim |E-E_c|^{z\nu} .$$

The simplest theory for understanding the effect of critical fluctuations close to a phase transition assumes that the order parameter (SDW, CDW, etc) couples locally with the relevant degree of freedom (spin, charge, etc). The resistivity is then given by the standard de Gennes-Friedel formula where the electron mean free path scales with the differential scattering cross section of the order parameter fluctuations.

In a classical phase transition - the behavior is driven by thermal fluctuations. The resistivity have the same kind of singularity as internal energy, implying that the critical behavior of the derivative of the resistivity is the same as the specific heat at the phase transition, implying that in a classical phase transition the critical behavior is marked by an inflection point in the resistivity at $T_c$.

Even though for a 2D system long range order is not possible at any finite temperature, the system can undergo a transition to quasi-long range order (KT transition) with the presence of vortex-antivortex pairs([24](#)). In this case the order parameter correlation length obeys the exponential dependence with temperature $T$:

$$\xi(T) \sim a \exp(b/|T-T_{KT}|^{1/2}),$$

where $a$ and $b$ are constants, and $T_{KT}(E)$ is the KT transition temperature which is a function of the external tuning parameter $E$. The resistivity scales with some power of the inverse correlation length and hence is supposed to have an exponential dependence with temperature.

## Semiconducting Group-VIB dichalcogenides

Because of the charge confinement and reduced dielectric screening, the optical properties of semiconducting 2D materials are dominated by excitonic effects. The optical spectra of $MoS_2$, one of the most studied TMDC, is characterised by three main transitions, named the A, B and C peaks. The A exciton is the lowest energy corresponding to the fundamental optical gap of the material. The corresponding exciton binding energy is ~1 eV according to theory. The B exciton also corresponds to a transition at the K point, but between the bands for opposite spin. The C peak is of a different nature, as it has contributions from excitons from a large, annular-shape region of the k-space with nearly identical transition energies. In nearly neutral monolayer samples, other quasiparticles have been observed, including positive and negatively charged excitons (i.e. trions) and biexcitons([25-27](#)). The large trion binding energies (20-30 meV) have no parallel in traditional semiconductors, and allow for these quasiparticles to be observed even at room temperature.

The series of Rydberg exciton states above the 1s (A) exciton of $WS_2$ reveal an exciton series that deviates significantly from the hydrogen model([28](#), [29](#)). Not only the 1s, 2s, 3s, … $n$s states have a closer spacing for small $n$, reflecting a weaker screening at short-range (~log$r$, where $r$ is the electron-hole separation)([28](#)), but also an entirely different dependence in the angular momentum. Ab-initio GW calculations show that the states in the same shell but with higher angular momentum are at lower energy levels, e.g. 3$d$, 3$p$ and 3$s$ are in increasing energy order.

From the technological point of view however, the most relevant transitions are those close to the fundamental gap at $K(K')$, which can be used for manipulating quantum information stored as spin and momentum (valley index) of individual electrons, holes, or excitons. The selection rule for optical transitions is valley dependent, with the $K(K')$ valley coupling exclusively to right (left) circular polarised light. Thus, the valley index, or pseudospin, can be controlled coherently using polarised light. Since the two valleys have non-zero and symmetrical Berry curvature, in the presence of in-plane electric field they give rise to Hall currents with sign depending on the valley index, an effect known as the valley Hall effect. The orbital magnetic moment is also valley dependent allowing for coupling with magnetic fields([30-36](#)).

TMDC quantum dots inherit the valley properties of the monolayer and therefore are appealing for valleytronics due to the possibility of controlling spin and valley states of single confined electrons or holes, for example via interaction with propagating single photons. Quantum dots can be created by growing finite islands on a monolayer substrate, or by applying confinement potentials using patterned electrodes.

## Phosphorene and group-IV monochalcogenides

Phosphorene, a monolayer of black phosphorus, is a monoelemental 2D material. Monolayer, few-layer, and bulk black phosphorus all are semiconducting, with a direct or nearly direct bandgap([4](#)). In addition to this, phosphorene has a very high mobility which can reach 1,000 $cm^2/V{\cdot}s$ for devices of thickness of about 10 nm at room temperatures([37](#)). This exceeds the carrier mobility of transition metal dichalcogenides. According to theoretical predictions, the phonon-limited hole mobilities can reach up to 10,000–26,000 $cm^2/V{\cdot}s$ for the monolayer (zigzag direction)([38](#)).

Both the optical and transport properties of phosphorene are highly anisotropic, as a consequence of its ortorhombic, wave-like structure. Optical selection rules dictate that the absorption threshold is lower for linear polarized light along the armchair direction than along the perpendicular direction; optical conductivity and Raman spectra are also anisotropic, providing a fast way to determine its lattice orientation.

Besides its optical and electronic properties, fundamental research in phosphorene has unraveled a growing number of physical phenomena including superconductivity, high thermoelectric figure of merit(*39*), birefringence, colossal UV absorption, etc.

Group-IV monochalcogenides SnS, GeS, SnSe and GeSe are isoelectronic with phosphorene and share its orthorhombic structure, but the two atom types break the inversion symmetry of the monolayer. As a consequence, they feature spin-orbit splitting (19-86 meV)(*40*) and piezoelectricity with large coupling between deformation and polarization change in plane (with piezoelectric coefficient $e_{33}$=7-23⨯$10^{-10}$ C/m, largely exceeding those of $MoS_2$ and hBN) (*41*).

SnS, SnSe and GeSe are semiconductors, with gap energies covering part of the infra-red and visible range for different numbers of layers(*40*). Even though the indirect bandgap (in most cases) makes them less attractive for optical applications, the existence of two pairs of two-fold degenerate valence and conduction band valleys, each placed on one of the principal axis of the Brillouin zone, makes them suitable for valleytronics applications. In this case, as the symmetry is orthorhombic, and therefore the valley manipulation processes are different from TMDCs. Valley pairs can be selected using linear rather than circularly polarized light. Secondly, there is no valley Hall effect, and therefore the transverse valley current under electric field is a second order effect.

Group-IV monochalcogenides are more stable against oxidation than phosphorene, can be grown by CVD, and have been recently exfoliated down to bilayer.

## Gallium and Indium monochalcogenides

GaX and InX (where X is a chacogen, like S, Se or Te) are yet another member of the family of hexagonal 2D materials. In this case each layer can be viewed as a double layer of metal M=Ga, In intercalated between two layers of chalcogen (X-M-M-X). The band structure of monolayers of such materials is rather unusual, having a 'Mexican hat' dispersion at the top of the valence band, leading to a high DoS(*42*, *43*) (bulk materials are most probably direct band-gap semiconductors). Thus, these materials have high and fast photoresponsibility(*44*, *45*), large second harmonic generation and therefore have attracted attention mostly due to their optical properties. In p-doped materials, if the Fermi level is close to this singularity, this gives rise to a ferromagnetic instability(*46*).

## Hexagonal boron nitride

hBN layers consist of hexagonal rings of alternating B and N atoms, with strong covalent $sp^2$ bonds and a lattice constant nearly identical to graphite. It is very resistant both to mechanical manipulation and chemical interactions. Moreover, it has a large bandgap in the UV range. For these reasons, hBN is a material of choice as encapsulating layer or substrate for 2D stacked devices, providing an atomically smooth surface free of dangling bonds and charge traps. hBN substrates leave the band structure of graphene near the Dirac point virtually unperturbed (if crystallographic orientations of the two crystals are misaligned) and dramatically improve the mobility of graphene devices(*47*, *48*).

## Oxide layers and other insulators

Many oxides have layered structures and can therefore be seen as a source for new 2D materials. These include lead oxide and lead salts (PbO, $Pb_2O(SO_4)$, $Na_2PbO_2$, etc.), phosphorus oxides and phosphates, molybdenum and vanadium oxides and other transition metal oxides. In these materials, the layers are often connected by weak covalent bonds, oxygen bridges or intercalating elements, and are normally non-stoichiometric (due to the presence of oxygen vacancies). Further, they are normally polycrystalline, and mechanical exfoliation methods are usually limited to those which are available in higher quality crystals. For the chemical means of production of such monolayers intercalation with bulky guest species such as tetrabutylammonium ions has been used. Some of these layered oxides have been studied due to their importance as battery cathode materials (e.g. $MoO_3$, $V_2O_5$ and other Mo and V oxides), superconductors (e.g. copper and cobalt layered oxides)([49](), [50]()), passivating layers (phosphorous oxide([51](), [52]())) and other areas of technological interest. Layered oxides allow for alloying, combination of different layers, and intercalation of ions and molecules, and therefore the possibilities of materials design are immense.

Amongst the most studied 2D insulators are hybrid perovskites, which are remarkable for their high optical absorption coefficient within the solar spectrum and strong luminescence. Thin-film perovskite-based solar cells have emerged with a 20% power conversion efficiency, a notable value for a new technology([53]()). A hybrid perovskite is formed by layers of a metal halide intercalated by layers of organic chains. The high solar cell efficiency is believed to be in great part due to the confinement of excitons to the layers. Few-layer hybrid perovskite have been isolated by mechanical exfoliation, and found to be stable in air in the timescale of minutes.

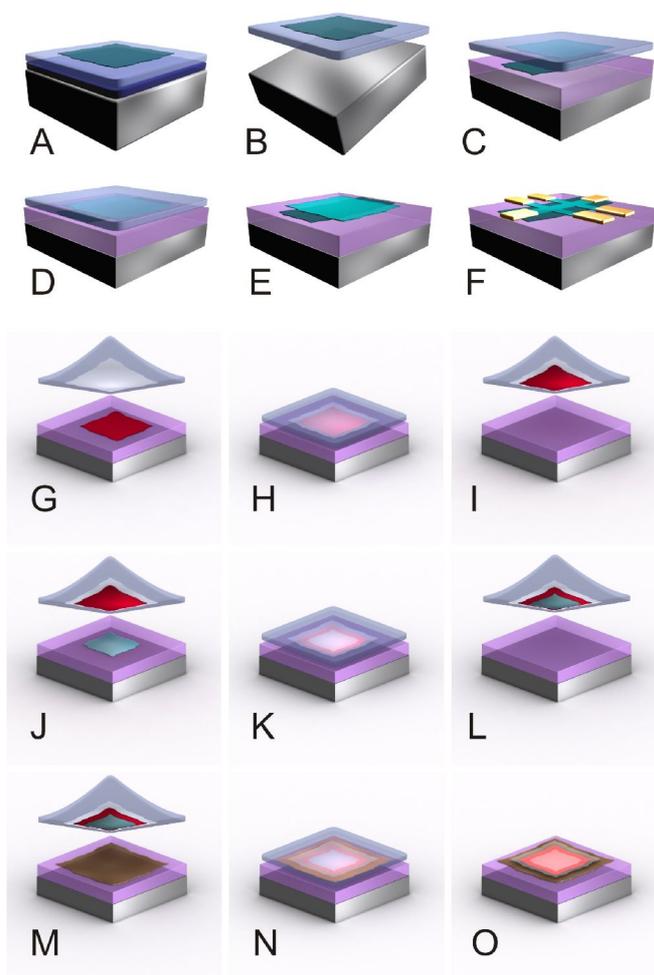

**Fig. 3 Wet transfer and "Pick and Lift" techniques for assembly of van der Waals heterostructures.** (**A-F**) Wet transfer technique. A 2D crystals prepared on a double sacrificial layer (A) is lifted on one of them by dissolving another (B). It is then aligned (C) and placed (D) on top of another 2D material. Upon the removal of the membrane (E) a set of contacts and mesa can be formed (F). The process could be repeated to add more layers on top. (**G-O**) "Pick and Lift" technique. A 2D crystals on a membrane (see (B)) is aligned (G) and then placed on top of another 2D crystal (H). Depending on the relative size of the two crystals it is then possible to lift both flakes on the same membrane (I). By repeating the process it is possible then to lift more crystals (J-L). Finally, the whole stack is placed on the crystal which will serve as the substrate (M-N) and the membrane is dissolved, exposing the whole stack (O).

## Novel van der Waals heterostructures

2D crystals can be assembled into heterostructures(*54*), where the monolayers are hold together by the van der Waals forces. Considering that a large number of 2D crystals is currently available, it should be possible to create a large variety of heterostructures. However, the assembling techniques, used at the moment (micromechanical stacking), only allow certain combinations of the interfaces. At the same time, the alternative technique, which potentially allows mass production of such structures - sequential growth of monolayers - comes with its own limitations and currently is at its infancy. Nevertheless, already now a large scope of novel experiments and prototypes has been done with van der Waals heterostructures, which makes it a versatile and practical tool for future experiments and applications.

## Assembly techniques

Currently, the most versatile technique for assembly of the heterostructures is the direct mechanical assembly. This technique flourished starting the revolutionary work of the Columbia group, which demonstrated the very high performance of graphene devices placed on hBN substrate(*47*).

The technique utilized in the early works is based on preparing a flake of 2D crystal (Fig. 3A-F) on a sacrificial membrane, aligning and placing it on another flake and then removing the membrane. The process is then repeated in order to deposit further layers. Although the crystals are exposed to sacrificial membrane and solvents, which can contaminate the interface, annealing allows one to remove the contaminants and achieve very high quality of the interfaces(*55*), reaching high mobility (~$10^6$ cm$^2$/V$\cdot$s) in graphene devices prepared this way.

A substantially cleaner method (dubbed "pick and lift") is based on strong van der Waals interaction which exists between many of such crystals. When the membrane with a 2D crystal on it is brought into contact with another 2D crystal, it is not being dissolved, but is lifted up, Fig. 3G-O. Then there is a chance that the second crystal will stick to the first and will be lifted together with it. The process can be repeated several times. This technique results in clean interfaces over large areas and yet higher electron mobility(*56*). Further advances could be achieved by transferring the whole process into a glovebox with controllable atmosphere.

## 1D contacts

The later method (Fig. 3G-O) has one particular disadvantage – having a completely assembled stack would prohibit one to make contacts to the inner layer. Luckily, it has been demonstrated that one can achieve various profiles of the edges of such a stack by reactive plasma etching. Thus, it is possible to etch the edge of the stack in such a way that the desired layer become exposed and can be contacted by metal evaporation(*56*), Fig. 4A. The contact resistance for graphene can be as low as 35 Ω·μm.

## Self-cleansing and self-cleaning mechanisms

The TEM studies(*55*) demonstrate that such interfaces can be atomically flat and free of any contamination, Fig. 4B-D (adopted from(*55*)). The reason for such behavior is the so called "self-cleansing" mechanism(*57*). If the affinity between the two 2D crystals is larger than the affinity between the crystals and the contaminants, then the energetically favorable situation is when the two crystals have the largest possible common interface. In order to achieve it, the contaminants are pushed away. This explains the observation of bubbles under transferred 2D crystals – those are the pockets of contamination pushed from the rest of the interface,

Fig. 4E-J (adapted from(*57*)). Such self-cleansing mechanism works only on certain pairs of crystals, Fig. 4E-G.

## Surface reconstruction

Potentially, the van der Waals interaction between two 2D crystals might lead to surface reconstruction. The most suitable candidates for the observation of such effects are crystals with similar lattice constants, such as graphene on hBN. The lattice constant of hBN is only 1.8% larger than that of graphene, which leads to the formation of moiré pattern(*5*).

It has been demonstrated that the most favorable configuration for graphene on hBN is when boron atoms lay on top of one of the sublattices in graphene and nitrogen is situated at the center of the hexagon(*58*). Graphene then tries to increase the area where the favorable configuration is achieved by stretching itself to match the interatomic spacing of hBN. Because of high Young modulus of graphene such perfect stacking cannot be achieved across the whole interface (unless the hBN can contract, as the loss in elastic energy would not be compensated by the gain in the van der Waals interaction). Thus, such stretching of graphene can only be local and the stretched regions would be separated by areas where graphene lattice is not commensurate with hBN, Fig 4K,L.

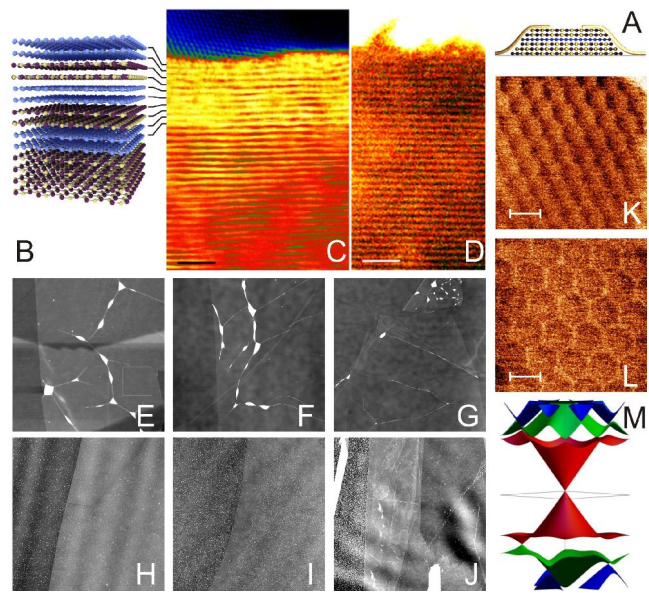

This effect was indeed observed for graphene on hBN when the crystallographic orientations of the two crystals are practically aligned(*10*). In this case the large regions of the moiré pattern where the two crystals are commensurate are separated by areas where graphene lattice is relaxed. This effect disappears when the graphene is misoriented with respect to hBN. Such commensurate-incommensurate transition happens at a critical angle which is given by the crystal mismatch(*10*).

Stacks of several other 2D crystals, including $MoS_2$ and $MoSe_2$ (*59*), $MoS_2$ and $WS_2$ (*60*), fluorographene and $MoS_2$ (*61*), and many others have been investigated for electronic properties(*62*) and possible surface reconstruction. Thus, layer-breathing phonon modes have been observed in Raman for $MoSe_2/MoS_2$ heterobilayers(*63*). However, because the lattice constant mismatch for those pairs is usually above 2%, the surface reconstruction would be hard to observe. It has been experimentally detected for silicene on $MoS_2$, where vertical buckling of silicene allows perfect stacking between the two crystals(*64*).

**Fig. 4 Morphology of the van der Waals heterostructures.** (**A**) 1D contacts to van der Waals heterostructures. Etching mesa in van der Waals heterostructures exposes the edges of the crystals inside the stack, which allows formation of 1D contacts. Here carbon atoms are represented by blue spheres; boron – yellow; nitrogen - brown. (**B-D**) TEM cross-section of graphene/hBN heterostructure. Scanning TEM image (C) of a structure schematically presented on (B). In (B) atom coloring same as in (A). (D) High-angle annular dark-field image of the same stack. Scare bars – 2nm. (**E-J**)AFM images of graphene transferred on other 2D crystals. Self-cleansing mechanism pushes contamination (hydrocarbons) away from graphene on hBN (E), $MoS_2$ (F) and $WS_2$ (G) interfaces, forcing the contamination to gather in bubbles. Instead, on the substrates with poor adhesion to graphene, such as mica (H), BSCCO (I), $V_2O_5$ (J), contamination is spread uniformly across the whole interface. The size of the images: 15 µm × 15 µm; z-scale: 4nm. (**K,L**) Local Young modulus for graphene on hBN for 3⁰ (K) and 0⁰ (L) misorientation angle. Note sharp domain walls in (L). Scale bars: 14nm. (**M**) Reconstructed electronic energy spectrum for graphene aligned on hBN.

## Spectrum reconstruction for graphene on hBN

Moiré pattern in graphene on hBN provides periodic scattering potential for electrons. This leads to the reconstruction of the electronic spectrum in graphene at the wavevectors determined by the periodicity of the moiré structure, as indeed has been observed in STM(*5*) and later in transport(*6-8*) and capacitance(*9*) measurements. Secondary Dirac points appear in the electronic spectrum both in the valence and in the conduction bands, Fig. 4M (adapted from(*65*)). The energy range where the spectrum is reconstructed is given by the strength of the van der Waals interaction between graphene and hBN and is estimated to be of the order of 50meV. Furthermore, the surface reconstruction leads to the strong asymmetry between the sublattices in graphene, which opens a gap in graphene spectrum.

## Capacitively coupled van der Waals heterostructures

Conceptually the simplest devices based on van der Waals heterostructures are those for capacitance measurements. hBN is an ideal insulator, which can sustain large electric fields (0.5V/layer and above), which allows preparation of capacitors with very thin dielectric. The use of thin dielectric ensures large contribution of the quantum capacitance, which is directly proportional to the DoS in the electrode, making capacitance measurements a viable tool to study both single particle and interaction phenomena in 2D materials. A number of systems have been investigated so far, including quantum capacitance in graphene(*66*), various sandwiches of graphene with TMDC(*57*) and black phosphorous(*67*).

Capacitive coupling between two graphene layers through a thin layer of hBN can also lead to a number of exciting phenomena. Thus, it allows for very high quality Coulomb drag devices, where two graphene layers, separated galvanically, interact through Coulomb forces between the charge carriers in the two layers(*68*). hBN, being atomically flat crystal with very large gap in the electronic spectrum allows very thin barriers (of the order of few nanometers) before any tunneling kicks in, bringing the two graphene layers closer than the characteristic distance between electrons in each of the layers (10nm for a characteristic density of $10^{12}$ cm$^{-2}$). This opens the new regime of effective zero layer separation in Coulomb drag experiments.

## Tunneling devices

Graphene can be combined with semiconductor and insulating 2D crystals to create tunnel junction(*69*). The use of hBN as a tunneling barrier is particularly attractive due to its large bandgap (~6eV), low number of impurity states within the barrier and high breakdown field. Since the position of the Fermi energy and the DoS in graphene can be varied by external gate, so as the tunneling current, which allows such structures to be used as field effect tunneling transistors (FETT)(*70*).

FETT architecture enables tunneling spectroscopy to probe DoS in graphene as well as to observe impurity- and phonon-assisted tunneling(*71*). Elastic tunneling through impurities gives peaks in *dI/dV* which positions depend on both bias and gate voltages (Fig. 5C). Inelastic phonon-assisted tunneling, from the other hand, is characterized by a set of plateaus in *dI/dV* independent of gate voltage(*71*) (more pronounced in $d^2I/dV_b^2$, Fig. 5B). When bias voltage is large enough to emit a phonon ($eV_b = \hbar\omega_{ph}$), an additional channel opens for electron tunneling, which increases transmission probability and, hence, tunnel conductance. Tunneling through impurities and with the phonon emission are especially well-visible if the crystallographic lattices of the two graphene electrodes are strongly misoriented with respect to each other, which prohibits direct electron tunneling since it is impossible to fulfil the momentum conservation requirements.

If the crystallographic lattices of the two graphene electrodes are aligned – momentum conservation for tunneling electrons can be achieved without impurity or phonon scattering. Rotational misalignment of the two graphene crystals corresponds to a relative rotation of the two graphene Brillouin zones in the reciprocal space. If the misalignment is small enough (< 2º) then the momentum difference between the electronic states in top and bottom graphene layers can be compensated electrostatically by applying bias and gate voltages(72), which leads to the resonant tunneling and observation of the negative differential resistance(72), Fig. 5D. Sharp negative differential resistance feature allows one to build a tunable radiofrequency oscillator with the potential to reach subterahertz frequencies.

Highest on/off ratio for such transistors can be achieved if the changes in the Fermi energy in graphene are comparable with the gap in the tunneling barrier – the situation achieved if hBN is replaced with $WS_2$ (on/off ratio $10^6$)(73) or $MoS_2$ (on/off ratio $10^3 - 10^4$ probably because of the presence of impurity bands)(70). Besides logic applications, tunneling in van der Waals heterostructures were exploited for memory devices(74) with a floating gate, logic circuits(75), radiofrequency oscillators(72), and resonant tunneling diodes(76).

## Interaction with light

Optoelectronic devices based on graphene(77) as well as on other 2D materials(78) have been studied intensively. However, graphene photodetectors typically have low responsivity, which is a consequence of low adsorption coefficient.

Such issues are eliminated when other 2D materials used for such purposes. Thus, transition metal dichalcogenides(78), GaS(79), InSe(80), black phosphorous(81) and other materials(82) have been used as photodetectors(83) in photodiode or photoconductor regimes. The advantages of using such materials is the large DoS (which guarantees large optical adsorption), their flexibility and the possibility of local gating, which allows creation of p-n junctions(84). Furthermore, often the bandgap in such materials depends on the number of layers(3), which allows one to control the spectral response in such devices.

## Van der Waals heterostructures for photovoltaic applications

Still, even larger opportunities open up when such materials are combined together. Thus, combinations of graphene (as a channel material) and transition metal dichalcogenides (as light sensitive material, where trapped charges are controlled by illumination) allow creation of simple and efficient phototransistors(85).

Combining together materials with different work-functions one can achieve that photoexcited electrons and holes are accumulated in different layers, giving rise to indirect excitons (as for example has been observed for the pairs $MoS_2/WSe_2$(86)) and $MoSe_2/WSe_2$(87) (Fig. 5E,F)). Such excitons typically have long lifetime and their binding energy could be controlled by controlling the distance between the semiconductor layers.

If p- and n-doped materials are used in such devices, then atomically sharp p-n junctions can be created(88, 89). Such devices are extremely efficient in carrier separation, so they demonstrate very high quantum efficiency (for instance, external quantum efficiency of above 60% was demonstrated for $GaTe/MoS_2$ devices(88)). Furthermore, their performance can be tuned externally by gate voltage, as, for instance, it has been demonstrated for black phosphorus/$MoS_2$ heterostructure(90).

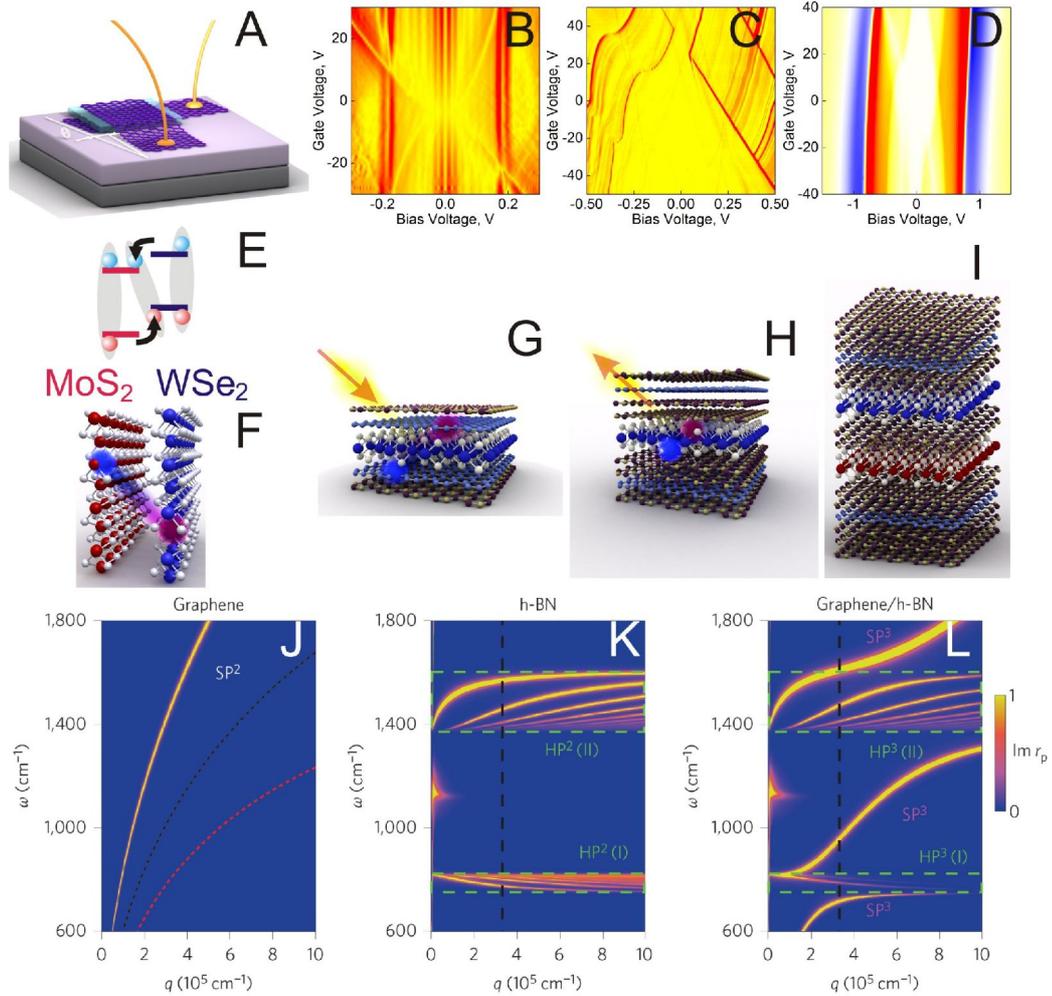

Fig. 5. **Electronic and optoelectronic applications of van der Waals heterostructures.** (**A-D**) Tunnelling in graphene-hBN-graphene tunnel transistors. (A) Schematic representation of graphene tunnelling device. Here graphene electrodes are dark-purple and hBN tunnelling barrier is light-blue. The electrodes can be aligned with respect to each other. (B) $d^2I/dV_b^2$ map of phonon-assisted tunnelling. Colour scale: yellow to red is 0 to $3.8 \cdot 10^{-5}$ $\Omega^{-1}V^{-1}$. (C) $dI/dV$ map of resonant tunnelling due to the presence of impurities in hBN tunnel layer. Colour scale: yellow to red is 0 to $7 \cdot 10^{-8}$ $\Omega^{-1}$. (D) $dI/dV$ map of resonant tunnelling with momentum conservation due to crystallographic alignment of two graphene electrodes. Blue colour marks the range of voltages where the negative differential conductivity is observed. Colour scale: blue to white to red is $-6 \cdot 10^{-6}$ to 0 to $6 \cdot 10^{-6}$ $\Omega^{-1}$. (**E,F**) Indirect excitons in $MoS_2$/$WSe_2$ heterostructure. Photoexcited electrons from $WSe_2$ are accumulated in $MoS_2$. Photoexcited holes from $MoS_2$ are accumulated in $WSe_2$. (**G**) TMDC (large blue and white spheres) sandwiched between two graphene (small light-blue spheres) electrodes for photovoltaic applications. Photocarriers generated in TMDC are separated into the opposite graphene electrodes due to electric field created by external gate (not shown). The structure can be encapsulated in hBN (purple and yellow spheres). (**H,I**) Vertical LED heterostructures. hBN barriers increase the dwell time of the electron and hole in the TMDC, allowing their radiative recombination. Multiple quantum wells, formed by different materials, can be utilised in such structures (I). Colouring of the atoms same as in (G). (**J-L**) Polaritonic dispersions of graphene, hBN and graphene/hBN heterostructure.

Even more efficient photovoltaic devices can be created by combining thin layers of transition metal dichalcogenides(*91*) or metal chalcogenides(*92*) with graphene. Sandwiching the photosensitive material between graphene electrodes one can achieve very efficient photocarriers extraction from the device into graphene electrodes (which typically form good ohmic contacts with the transition metal dichalcogenides and serves as a transparent electrode as well). Since these structures are typically symmetric (Fig. 5G), one needs to create electric field inside the TMDC to produce efficient carrier separation by either bias, external gate

(because electric field is not fully screened by graphene due to its low DoS) or by different doping of the two graphene layers.

## Light emitting diodes

The p-n junctions described above can be operated in the regime of electrical injection of the charge carriers, which leads to electron-hole recombination and light emission(*89*). Such arrangement, however, suffer from the requirements of synthesizing p- and n-type materials, which has not been demonstrated for all the 2D crystals yet, and from the fact that the resistance of the junction is comparable with the resistances of the p- and n-electrodes, which makes it hard to control the current distribution.

More straightforward arrangement is the carrier injection from highly conductive transparent electrodes directly into the 2D material in a vertical structure. Such scheme, however, requires careful control of the dwell time of the injected electrons and holes in the semiconductor crystal, since the photoemission is a slow process in comparison with the characteristic time required to penetrate the junction between graphene and the semiconductor. The dwell time can be controlled by introducing additional tunnel barriers (Fig. 5H). Thus, 2-3 layers of hBN have been used(*93*) in order to increase the time electrons and holes spend inside the monolayer TMDC, allowing their radiative recombination. Especially efficient are devices based on $WSe_2$, which quantum efficiency increases with increasing both temperature and the injection current, reaching 20% at room temperature(*94*). One can increase the quantum efficiency of such structures by placing several layers of TMDC in series(*93*) (Fig. 5I).

## Plasmonic devices

Plasmons in graphene attract a lot of attention due to a possibility of tuning their frequency by changing the carrier concentration, and thus the plasmonic frequency(*95*). Simultaneously, plasmonic and phonon-polaritonic properties have been studied in other 2D materials: thus hBN, being polar dielectric, supports surface phonon polaritons with very low optical losses(*96*).

A number of new polaritonic effects can be seen in van der Waals heterostructures. Thus, encapsulation of graphene with hBN allows one to eliminate the scattering of graphene plasmons with impurities, increasing their inverse damping ratio by a factor of 5 in comparison with bare graphene(*97*). By sandwiching several graphene layers separated by hBN spacers one can hybridise plasmonic modes in such multilayers, which can be further controlled with external gate voltage(*98*).

In such heterostructures it is possible to enter a regime where the plasmon polaritons in graphene and the phonon polaritons in hBN coexist, Fig. 5J-L (adapted from(*99*)). Strong coupling between the two leads to formation of the new collective modes – plasmon-phonon polaritons(*100*). Both the amplitude and the wavelength of the new collective modes can be controlled by gating graphene.

In aligned graphene/hBN heterostructures the formation of the moiré pattern provides further modification of the graphene plasmonic spectrum. Zone folding results in the formation of the secondary Dirac points(*5-8*, *65*), Fig. 4M, which allows new type of vertical optical transitions. Such transitions are immediately reflected in the modified damping factor, which exhibits maximum at such energies(*101*). It has also been predicted that new plasmonic modes with carrier density dependence characteristic to parabolic electronic bands should appear in the vicinity of the van Hove singularities of the reconstructed spectrum(*101*) (Fig. 4M).

## Assembling van der Waals heterostructures in liquid and from liquid phase exfoliated 2D materials

A very powerful method of preparing graphene, which can also be extended to other materials is liquid phase exfoliation(*102*). Ink formulation, based on such suspensions, gave rise to development of graphene-based printed electronics(*103*). However, many applications would strongly benefit from the properties which are beyond the capabilities of graphene inks. Thus, high thermal conductivity combined with dielectric property can be delivered by hBN and optoelectronic capabilities can be delivered by inks of 2D semiconductors.

The ability to print combinations of such materials opens the door to low-cost fabrication of various devices(*104*). Planar(*105*) and vertical(*106*) photovoltaic devices based on TMDC, as well as planar(*107*) and tunneling transistors(*106*) based on graphene and hBN have recently been demonstrated.

By solution synthesis of the 2D crystals or by controlling the charge on individual flakes in the suspensions, heterostructures can be formed directly in the liquid phase(*108*) and can be used for energy applications. Thus, $MoSe_2$/graphene structures have been used for Li ion batteries applications(*109*). Such structures have been used for catalytic applications(*110*).

## Growing van der Waals heterostructures

Direct growth methods such as chemical vapor deposition (CVD) are promising techniques for scalable manufacturing of van der Waals heterostructures(*111*). Such techniques can be grouped as following: (i) sequential CVD growth of 2D crystal on top of mechanically transferred or grown 2D material, (ii) direct growth of TMDC heterostructures by vapor-solid reactions, and (iii) van der Waals epitaxy. State-of-the-art CVD, direct growth and van der Waals epitaxy methods have already enabled growing many vertical heterostructures: graphene/hBN(*112-116*), $MoS_2$/graphene(*117-120*), GaSe/graphene(*121*), $MoS_2$/hBN(*122, 123*), $WS_2$/hBN(*124*), $MoTe_2$/$MoS_2$(*125*), $WS_2$/$MoS_2$(*126*), $VSe_2$/$GeSe_2$(*127*), $MoSe_2$/$Bi_2Se_3$(*128*), $MoSe_2$/$HfSe_2$(*129*), $MoS_2$/$WSe_2$/graphene and $WSe_2$/$MoSe_2$/graphene(*76*), to list a few.

An important achievement was *in situ* CVD growth of encapsulated graphene in hBN/graphene/hBN heterostructure, which has demonstrated the scalability of graphene-based high mobility field effect transistors(*116*). Also, some of the TMDC heterostructures can be grown directly in a single step process: $WS_2$/$MoS_2$ heterobilayer was grown on $SiO_2$/Si substrate at 850ºC from precursors (W, S, $MoO_3$) placed in the growth tube (Fig. 6A, adapted from (*126*)). Because of the difference in the growth rates of $MoS_2$ and $WS_2$, the formation of $Mo_xW_{1-x}S_2$ alloy is suppressed. Clean interface enabled a band alignment of the two constituent layers which led to an observation of indirect excitons in $WS_2$/$MoS_2$ heterostructure.

## Van der Waals epitaxy

Van der Waals epitaxy was introduced more than 30 years ago with the growth of a $NbSe_2$ monolayer on a cleaved face of $MoS_2$ bulk crystal (*130*). This work also led to a successful growth of monolayer $MoSe_2$ on SnS₂ (*131*), and even of a three-component heterostructure of monolayer $NbSe_2$/trilayer $MoSe_2$/mica (*132*).

To grow graphene on hBN one can use plasma to break down precursor methane molecules(*48*), and the growth occurred at 500ºC during 2-3h on hBN crystals mechanically exfoliated on $SiO_2$/Si substrate. Van der Waals interactions during epitaxial growth defined the preferential growth directions so that graphene crystals were aligned to hBN substrate, Fig. 6E-G (adapted from (*48*)).

Mechanically exfoliated hBN has also served as a substrate for CVD-based van der Waals epitaxy of a rotationally commensurate MoS$_2$/hBN heterostructure (*122*). Another example of van der Waals epitaxy is the growth of high-quality wafer-scale MoSe$_2$/Bi$_2$Se$_3$ heterostructures on a low-cost dielectric substrate AlN/Si in ultrahigh vacuum conditions (*128*).

Graphene is also a good substrate for van der Waals epitaxy – grown WSe$_2$/graphene heterostructures show atomically sharp interface and nearly perfect crystallographic orientation between graphene and WSe$_2$ despite of a large (23%) lattice mismatch (*133*). A few-layer MoS$_2$ and hBN were also grown using epitaxial graphene as a growth substrate(*117*, *134*). Recently monolayers of WSe$_2$ and MoS$_2$ were grown on free-standing CVD graphene(*135*). TMDC crystals were also explored as substrates for epitaxy when MoTe$_2$ monolayer was grown on a bulk MoS$_2$ substrate (*125*).

Finally, van der Waals epitaxy can be repeated several times to grow complex multicomponent heterostructures, such as atomically thin resonant tunneling diode based on MoS$_2$/WSe$_2$/graphene and WSe$_2$/MoSe$_2$/graphene, see Fig. 6H-L (adapted from (*76*)). To this end epitaxial graphene trilayer was used as a substrate to grow monolayers of either MoS$_2$ (at 750ºC) or WSe$_2$ (at 850ºC) via powder vaporization or metal organic CVD processes. After that, a second TMDC layer (WSe$_2$ or MoS$_2$) was grown on top of the grown heterostructure.

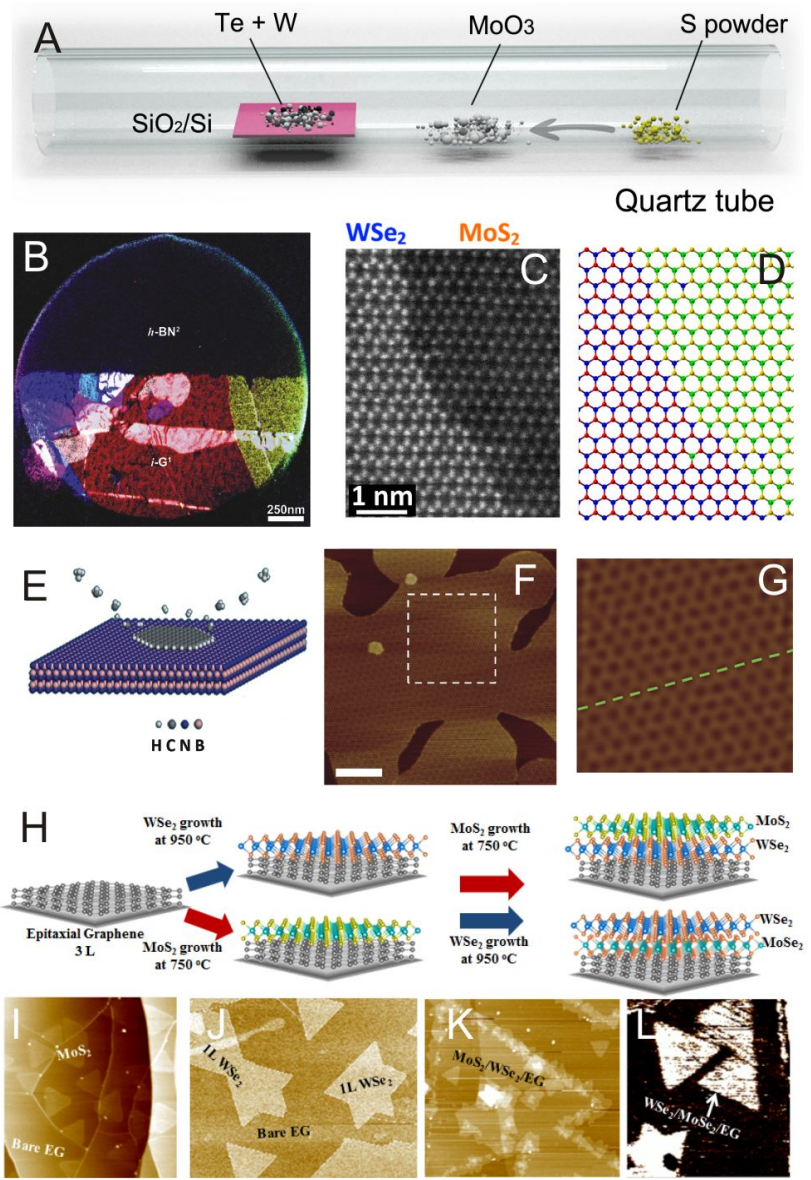

**Fig. 6. Van der Waals epitaxy of vertical and in-plane heterostructures.** (**A**) Schematic of the growth process of vertically stacked and in-plane WS$_2$/MoS$_2$ heterostructures. (**B**) False-colour DF-TEM image of a suspended hBN-graphene in-plane heterostructure. (**C,D**) High-resolution STEM image and an atomic model of WSe$_2$-MoS$_2$ in-plane heterostructure. (**E**) Schematics of vertical van der Waals heteroepitaxy of graphene on hBN. (**F,G**) Moiré pattern of graphene/hBN heterostructure as observed in tapping-mode AFM, and in high-pass-filtered inverse FFT of the dashed square region in F. (**H**) Schematics of a growth of MoS$_2$/WSe$_2$/graphene and WSe$_2$/MoSe$_2$/graphene. Synthesis of both 3-component heterostructures begins by growing 3 layers epitaxial graphene, followed by MOCVD growth of either MoS$_2$ (I) or WSe$_2$ (J). Then, another TMDC layer is grown by vapour transfer of either MoS$_2$ (K) or WSe$_2$ (L). (**I-K**) AFM images of MoS$_2$/graphene, WSe$_2$/graphene, and MoS$_2$/WSe$_2$/graphene vertical heterostructures, respectively. (**L**) Conductive AFM image of WSe$_2$/MoSe$_2$/graphene heterostructure, colour scale: black-to-white is 0-100 pA. Note that due to Se-S ion exchange, a layer of MoSe$_2$ forms from the original MoS$_2$ layer.

## Lateral heterostructures

Lateral heterostructures can also be grown by a variety of methods. Thus, CVD-grown graphene was lithographically patterned and etched away and hBN was grown via CVD, forming lateral 1D heterojunctions (Fig. 6B, adapted from(*136*)). Beyond graphene and hBN, lateral heterostructures based on 2D TMDCs can be disruptive for integrated optoelectronic devices. Although direct growth favors TMDC alloys because of a similar chemistry and a small lattice mismatch between different TMDCs(*137*), a two-step epitaxial growth of $MoS_2$-$WSe_2$ lateral heterostructure was recently demonstrated, Fig. 6C,D (adapted from(*138*)). To avoid alloying during growth, two separate temperature regimes were used(*138*). Atomically sharp $WSe_2$-$MoS_2$ heterojunction has a depletion width of ≈300nm due to the potential difference between $MoS_2$ and $WSe_2$ regions.

Lateral heterostructures of $MoS_2$-$WS_2$ and $WSe_2$-$MoSe_2$ were grown directly by controlling the growth temperature at ~650ºC(*126*). The growth at relatively low temperatures was facilitated by either introducing tellurium into CVD process (*126*) or via using perylene-based growth promotors(*139*). The use of growth-promoting perylene-based aromatic molecules was recently extended to stitch together largely dissimilar 2D materials.

## Conclusion

The family of 2D crystals grows both in terms of variety and number of materials, and it looks like this process is only at the beginning. Almost every new member brings excitement in terms of the unusual physics it possesses. The 2D physics in such materials (such as KT transitions) is only starting to emerge. Still, we would like to argue that the most interesting phenomena can be realized in van der Waals heterostructures, which now can not only be mechanically assembled, but also grown by a variety of techniques. Among the unsolved problems is the control of the surface reconstruction, charge transfers and build-in electric fields in such heterostructures. The standard band diagrams with quasi-electric fields is not a useful concept in 2D heterostructures, and a new framework has to be developed.

This work was supported by the EU FP7 Graphene Flagship Project 604391, ERC Synergy Grant, Hetero2D, EPSRC (Towards Engineering Grand Challenges and Fellowship programs), the Royal Society, US Army Research Office, US Navy Research Office and US Airforce Research Office. AHCN and ASC acknowledge the National Research Foundation of Singapore under the Prime Minister's Office for Financial support under the Mid Size Centre Grant. The authors are grateful to Andre K. Geim, Irina V. Grigorieva, Vladimir I. Fal'ko, Roman V. Gorbachev, Freddie Withers, Andy V. Kretinin, Conrad Gesner, Sarah J. Haigh.